\documentstyle[prl,aps,twocolumn,epsfig]{revtex}

\begin{document}

\twocolumn[\hsize\textwidth\columnwidth\hsize\csname @twocolumnfalse\endcsname

\title{Diffuse Neutron Scattering Study of a Disordered Complex Perovskite Pb(Zn$_{1/3}$Nb$_{2/3}$)O$_{3}$ Crystal}
\author{D. La-Orauttapong,$^{1}$ J. Toulouse,$^{1}$  J. L. Robertson,$^{2}$  and Z.-G. Ye$^{3}$}
\address{$^{1}$Physics Department, Lehigh University, Bethlehem, PA 18015-3182}
\address{$^{2}$Oak Ridge National Laboratory, Solid State Division, Oak Ridge, TN 37831-6393}
\address{$^{3}$Department of Chemistry, Simon Fraser University, Burnaby, BC, V5A 1S6, Canada}
\maketitle

\begin{abstract}
Diffuse scattering around the (110) reciprocal lattice point has been
investigated by elastic neutron scattering in the paraelectric and the
relaxor phases of the disordered complex perovskite crystal-Pb(Zn$_{1/3}$Nb$_{2/3}$)O$_{3}$\ (PZN). \ 
The appearance of a diffuse intensity peak
indicates the formation of polar nanoregions at $\ $temperature $T^{\ast }$,
approximately 40K above $T_{c}$=413K. \ The analysis of this diffuse
scattering indicates that these regions are in the shape of ellipsoids, more
extended in the $<111>$ direction than in the $<001>$ direction. \ The
quantitative analysis provides an estimate of the correlation length, $\xi $, 
or size of the regions and shows that $\xi _{\left< 111\right> }\sim
1.2\xi _{\left< 001\right> }$, consistent with the primary or dominant
displacement of Pb leading to the low temperature rhombohedral phase. \ Both
the appearance of the polar regions at $T^{\ast }$ and the structural
transition at $T_{c}$ are marked by kinks in the $\xi _{\left< 111\right> }$
curve but not in the $\xi _{\left< 001\right> }$ one, also indicating that
the primary changes take place in a $<111>$ direction at both temperatures.

\end{abstract}
\vskip1pc]

\narrowtext

Many of the relaxor ferroelectrics known today are lead-based compounds with
the perovskite structure. \ In addition to the characteristic frequency
dispersion of their dielectric constant, several of them exhibit remarkable
piezoelectric or electrostrictive properties that are
finding important applications, e.g. as transducers and actuators. \ The Pb$%
^{2+}$-containing relaxor perovskites such as Pb$($R$_{1/3}$Nb$_{2/3})$O$%
_{3} $ (R=Mg$^{2+}$, Zn$^{2+}$) have a common ABO$_{3}$ cubic perovskite
structure in which the B-site can be occupied by $\frac{1}{3}$R$^{2+}$ and 
$\frac{2}{3}$Nb$^{5+}$. \ Because of the different atomic radii and valences 
of the B-site cation, PMN and PZN exhibit short-range chemical ordering\cite
{Yokomizo et al_1278_1970,Randall_Bhalla_327_1990}. \ Over the years, 
these relaxors and their properties have been described in a variety of ways, 
most often in terms of the formation of polar micro- or nanoregions 
\cite{Smolensky_26_1970,Cross_241_1987}. \ However, these models are primarily
based on indirect experimental evidence for such regions and more direct evidence is necessary \ 
in order to elucidate the true origin of the relaxor behavior. For this purpose, neutron and
X-ray techniques are most suitable, as they can provide evidence of local
structural ordering. \ Obtaining such evidence is crucial if the formation
of polar regions is indeed responsible for the relaxor behavior. \ 

PZN is a prototype relaxor ferroelectric. \ 
It was found to exhibit a large dielectric dispersion and a broad dielectric maximum that
depended on both frequency and temperature. \ Earlier studies
also reported a structural phase transition from cubic to rhombohedral symmetry 
near 413K\cite{Khuchua et al_194_1968,Kuwata et al_863_1979}, which falls in 
the temperature region of the maximum of the dielectric peak. \ More recent 
studies only mention the coexistence of cubic and rhombohedral phases, with 
polar nanodomains growing into polar microdomains \cite{Mulvihill et al_1462_1997} 
and their volume fraction increasing with decreasing temperature. \ Strain appears 
to play an important role in a nanodomain-to-microdomain phase transition, which
therefore resembles a martensitic phase transformation\cite{Mulvihill et
al_1462_1997}. \ In the last few years, PZN has also been intensively
investigated in the form of a solid solution with PbTiO$_{3}$\cite
{Kuwata et al_1298_1982,Liu et al_2810_1999,Gehring et al_5216_2000}, 
because of its enhanced dielectric and piezoelectric properties. \
PZN is already a disordered perovskite in the cubic phase\cite{Iwase et al_1419_1999}, 
with individual Pb atoms shifted from the A-site position along one of eight 
equivalent $\left\langle 111\right\rangle $ directions and O atoms forming a \ 
ring-like distribution within $\left\{ 100\right\} $ planes. \ 
In contrast, the Zn/Nb atoms remain located at the normal B-site position. \ 
In the rhombohedral phase, the displacements of Pb and O
atoms become correlated, both along a specific [111] direction, with $\delta r_{\text{%
Pb}}$ = 0.31 \AA\ and $\delta r_{\text{O}}$ = 0.17 \AA\ at 295K. \ More
specifically, a recent X-ray study \cite{Takesue et al_2001} of the related
system PMN suggests that the three oxygen atoms may be displaced along
three converging [110] directions, all three pointing towards the same A
position, such that the average oxygen displacement or the net oxygen dipole 
moment in the unit cell is indeed along the same [111] direction as that of Pb. \ 
It is important to note that the displacement of Pb \cite{Iwase et al_1419_1999} 
as well as its neutron scattering length are almost twice those of the O atom. \ 

In the present paper, we report the results of a diffuse neutron scattering
study of PZN, in order to identify and directly monitor the formation of
polar nanoregions resulting from the correlated displacements of the Pb and
O atoms. \ This study follows a similar study of another mixed perovskite
relaxor, K$_{1-x}$Li$_{x}$TaO$_{3}$ (KLT), which provided the correlation
length or size of the polar nanoregions as a function of temperature \cite
{Yong et al_14736_2000}. \ In contrast to KLT, PZN exhibits not only
substitutional chemical disorder but also charge disorder. \ Moreover, due
to the lone electron pair effects, Pb$^{2+}$ions form partially covalent
bonds with the neighboring oxygen atoms, which results in the off-centering
of lead and in an additional local electric dipole moment \cite{Egami et
al_S935_1998}. \ Our purpose here is to provide direct evidence for these
regions and to characterize them in a single crystal of PZN.

In complex perovskites such as PZN, the diffuse scattering has two possible
origins \cite{Krivoglaz_NewYork_1969}: i) different scattering factors of
the two possible atoms (B$_{1}$,B$_{2}$) occupying a given B site ($j$), or
ii) static displacements away from this site. \ Both factors arise from
concentration inhomogeneities in the\ compound and may simultaneously
contribute to the diffuse scattering. \ It is well known that the static 
structure factor is given by \cite{Schmatz_NewYork_London_1973}:

\begin{equation}
F(\text{{\bf Q}})={\sum_{j}}b_{j}e^{i\text{{\bf Q{.r}}}_{j}}e^{-W_{j}}
\end{equation}

\noindent where $b_{j}$\ is the neutron scattering length, $e^{-W_{j}}$ is
the Debye-Waller factor for the $j$th atom with coordinate ${\bf r}_{j}$ in
the unit cell and the summation is carried out over all positions occupied
in the unit cell. \ The scattering vector {\bf Q} is defined as 
${\bf k}-{\bf k}^{{\acute{}}}$. \ In the presence case, small displacements, 
${\bf \delta r}$, of the atoms from 
their normal positions distort the local cubic symmetry such that \ 
${\bf r}_{j}$=$({\bf r}+ {\bf \delta r})_{j}$. At high temperature, these displacements 
vary both in magnitude and direction. \ The scattered intensity is proportional to 
$F^{2}=F.F^{\ast }$ so that, experimentally, it is possible to determine the
amplitude $F$ of the structure factor but not its phase. \ For the sake of
argument we omit the Debye-Waller factor in the following calculation. \ In accordance with
Ref.\cite{Iwase et al_1419_1999}, which reported the Zn/Nb atoms at their
ideal perovskite position at all temperatures, no displacement is included
in eq. (1) for the atoms at the B-site. With the origin chosen at the Pb site, eq. (1) \ 
and for small displacements of the Pb and O atoms, the neutron scattering structure factor $F$ can then be evaluated at
the (110) reciprocal lattice point:

\begin{eqnarray}
F(\text{{\bf Q}})&=&[{\underbrace{b_{\text{Pb}}+b_{\text{Zn/Nb}}-b_{\text{O }%
}}_{\text{Bragg}}}]  \nonumber \\
& &+ 
\begin{array}{c}
{\underbrace{i\text{{\bf Q.}}\left[ \delta \text{{\bf {r}}}_{\text{Pb}}b_{%
\text{Pb}}+\left( \delta \text{{\bf {r}}}_{\text{O}1}-\delta \text{{\bf {r}}}%
_{\text{O}2}-\delta \text{{\bf {r}}}_{\text{O}3}\right) b_{\text{O}}\right] }%
_{\text{Diffuse}}}
\end{array}
\end{eqnarray}

\noindent where $b_{\text{Pb}}$, $b_{\text{Zn/Nb}}$, and $b_{\text{O}}$ are
the scattering lengths of Pb, Zn/Nb, and O atoms respectively. \ The
scattering lengths for the Zn/Nb atoms can be found from $b_{\text{Zn/Nb}}=%
\frac{1}{3}b_{\text{Zn}}$ +$\frac{2}{3}b_{\text{Nb}}$. \ In the structure factor above, 
we have separated the Bragg scattering
contribution which contains information about the average lattice from the
diffuse scattering which arises from local deviations from the ideal
structure. \ These deviations result from correlated displacements of atoms. \ 
In the presence case, the diffuse scattering is primarily associated with 
the Pb atom, which is strongly off-centered in the PbO$_{12}$ unit and has 
almost twice the neutron scattering length of the O atom. \ In addition, because \ 
of the Pb displacement along the [111] direction, ${\bf \delta }r_{\text{Pb}}$is \ 
expected to make a greater contribution in that direction than along the [001] and [110] \ 
directions. \ 

The neutron scattering experiments were performed on the HB-1 triple-axis
spectrometer of the High Flux Isotope Reactor at Oak Ridge National
Laboratory. \ Pyrolytic graphite (002) crystals were used for both the monochromator and
analyzer, to suppress harmonic contamination. \ The collimation was 
48$^{{\acute{}}}$-20$^{{\acute{}}}$-20$^{{\acute{}}}$-70$^{{\acute{}}}$ 
and the incident neutron energy was 13.6 meV ($\lambda $=2.453\AA ). \
Initially, the scattering around the (110) reciprocal lattice point in the [110]-[001] 
scattering plane was studied because the (110) reflection has the largest
structure factor \cite{Yokomizo et al_1278_1970}.  In each scan, the stepsize was 
$q=$0.01 reciprocal lattice units (rlu) or 0.016 \AA $^{-1}$. \ 
The yellowish single crystal of Pb(Zn$_{1/3}$Nb$_{2/3}$)O$_{3}$
used in the experiment was grown by a high temperature solution growth
technique \cite{Zhang et al_96_2000}. \ The measurements were performed upon
cooling from 550 to 295K and no external electric field was applied.
\begin{figure}[tbp]
\epsfig{width=0.8 \linewidth, figure=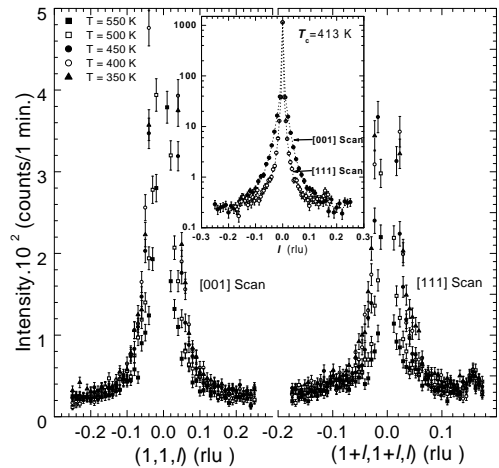 }
\caption{Elastic diffuse scattering of neutrons in the [001] and [111]
directions near the (110) reciprocal lattice point at T=550, 500, 450, 400,
and 350K. Inset is scattering intensities at $T_{c}$=413K in the [001]and [111] directions.}
\end{figure}
The narrow Bragg peak as well as the relatively broad diffuse scattering 
peak at the (110)reciprocal lattice point are presented in Fig.1 for several temperatures. \ 
As expected, the Bragg scattering is equally strong in both the [001] and [111]
scans(inset of Fig.1). \ However, the diffuse scattering is more extended in the [001] 
than in the [111] direction, having the form of a
rod along a cubic direction. \ It is important to note that the diffuse
scattering is more extended in a direction in which the correlation length $%
\xi $ is shorter. \ We will discuss this point later. \ In order to separate the Bragg 
and diffuse scattering contributions, the intensity, $I(q)$, was fitted as the sum \ 
of a Gaussian and a Lorentzian, using a least-squares method(Fig.2). \ 
In the inset of Fig.2, the Bragg intensity is seen to increase slowly, starting at \ 
approximately 500K, then more rapidly below 400K.  The latter rapid increase \ 
of the Bragg intensity can be explained by i) a relief
of extinction due to an increase in mosaicity below the transition or ii) an
unusual behavior of the Debye-Waller factor due to freezing of certain
atomic motion at lower temperatures.
\ A broadening of the Bragg peak is normally observed to accompany the
freezing process but is not observed here. \ This would rule out the effect
of the Debye-Waller factor in the present case and favor the first
explanation. \ In fact, a similar behavior was observed in a single crystal
of PMN by Gosula et al. \cite{Gosula et al_221_2000} using X-ray diffuse
scattering. \ They speculated that the sharp increase in intensity arose
from the anti-ferrodistortive ordering that developed at low temperatures. \
This anti-ferrodistortive ordering grows to a few hundred angstroms with
decreasing temperature but never reaches truly long-range order.
\begin{figure}[tbp]
\epsfig{width=0.8 \linewidth, figure=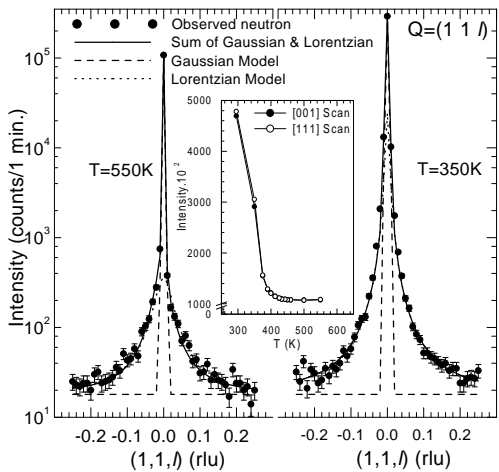 }
\caption{Elastic neutron scattering in the [001] direction at T=550 and 350K. 
The solid curve shows the scattering profile fitted by the sum of a Gaussian and a
Lorentzian and the two components (dash and dot curves)are shown separately. 
Inset is temperature dependence of Bragg peak intensities in the [001] and [111] directions.}
\end{figure}
Turning to the diffuse scattering component, in Fig.2, the $q$ or $l$%
-dependence of the scattered intensity is presented on a semilogarithmic
plot at 550K and 350K. \ At 550 K the diffuse scattering intensity is slightly 
asymmetric which could be related to rather large deformations of the crystal lattice 
giving rise to Huang scattering\cite{Krivoglaz_NewYork_1969}. \ The Lorentzian function 
used in the fit of the experimental $I(q)$ curves, yielded a peak intensity and 
a full width at half- maximum $\left( \Delta q_{_{FWHM}}\right) $. \ Such a Lorentzian
lineshape is predicted by the Ornstein-Zernike model \cite{Stanley_1971}:
$I\simeq \frac{1}{q^{2}+\xi ^{-2}}$ where $q=l$ is the momentum transfer relative to the 
{\bf Q}=110 Bragg reflection and $\xi $ is the correlation 
length. \ The Ornstein-Zernike model assumes a correlation function of the form 
$\frac{e^{-\frac{r}{\xi }}}{r}$ for the polarization. \ Therefore the diffuse 
scattering width at half maximum, $\Delta q_{_{FWHM}}$, provides an estimate of $\xi $ or,
equivalently, of the size of the polar nanoregions:$\xi =\frac{2}{\Delta q_{_{FWHM}}}$ 
where $a$ is the lattice parameter. \ The Lorentzian fit parameters, 
$I\left( q\right) _{\max }$, $\Delta q_{_{FWHM}}^{2}$, and $\xi $, are shown in Fig.3 (a) 
and (b) as a function of temperature. \ In both the $[001]$, and $[111]$ directions, 
the results indicate that:
\begin{figure}[tbp]
\epsfig{width=0.8 \linewidth, figure=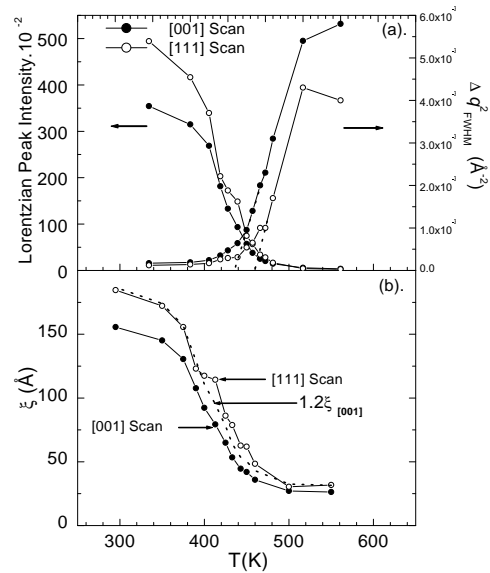 }
\caption{Temperature dependence of (a).the Lorentzian Peak Intensity and
square of the FWHM,$\Delta q_{_{FWHM}}^{2}$, (b).the
correlation length,$\protect\xi$ in the [001] and [111] directions.}
\end{figure}
i) the diffuse scattering peak intensity, $I\left( q\right) _{\max }$,
increases as the temperature decreases. \ It is interesting however to note
that the peak intensity, as measured in the $[111]$ direction, goes through
a plateau around the transition temperature $(T_{c}=413$K$)$ before
increasing again below 400K. \ This feature is much less visible in the $%
[001]$ direction, for which the intensity increases continuously. \ Fig.3
(a) also shows that the strongest diffuse scattering is observed in the $%
[111]$ direction, an observation that is consistent with the displacement
of the Pb atoms producing the local polarization in the
rhombohedral phase. \ The diffuse scattering intensity is in fact generally
proportional to $\left\langle P_{local}^{2}\right\rangle $ \cite{Nomura et
al_39_1999}.\ \ Consequently, an increase of the diffuse scattering
intensity with decreasing temperature reflects the growth of the polar
clusters or the net correlated displacements of lead and oxygen atoms in a $%
[111]$ direction as reported in Ref.\cite{Iwase et al_1419_1999} for PZN. \
It is also in agreement with the structure factor calculation $F$ developed
above, which predicts a greater magnitude of $F.F^{\star \text{ }}$along the 
$[111]$ direction than along the others.

ii) the width squared, $\Delta q_{_{FWHM}}^{2}$, initially decreases almost
linearly with temperature, but then deviates from a straight line at
approximately 450K, or 40 degrees above the transition $(T_{c}=413$K$)$ and
subsequently levels off at low temperature.\ The initial linear dependence
is\ the manifestation of a Curie-Weiss law. \ This should be expected in a
temperature range in which the polarization fluctuations are dynamic, since
the correlation length squared, $\xi ^{2}$, is proportional to the
dielectric constant, $\epsilon $, \cite
{Jona_Shirane_NewYorK_1962} which itself is expected to follow a Curie-Weiss 
law initially:

\begin{equation}
\frac{1}{\Delta q_{_{FWHM}}^{2}}\sim \epsilon =\frac{C}{T-T_{c}}
\end{equation}

\noindent where $C$ is the Curie constant. \ The departure of $\Delta
q_{_{FWHM}}^{2}$from a linear dependence, seen in Fig.3 (a), marks the
appearance of long-lived polar fluctuations or permanent polar regions in
the crystal, accompanied by local strain fields. \ This also explains the
simultaneous increase in Bragg intensity noted above, due to an increase in
mosaicity. \ It is worth noting that a very similar evolution has been
reported by Iwase et al. \cite{Iwase et al_1419_1999} for the lattice
constant, but with a departure from a linear dependence at approximately
550K instead of 450K. \ However, their measurements were made on a powder
obtained from a ground crystal and it is well known that relaxors are very
susceptible to internal strains. \ The correlation length, $\xi $, shown in Fig.3 (b)
increases continuously with decreasing temperature and saturates at low
temperatures. \ This is uncommon for conventional ferroelectrics, in which the
correlation length decreases below the transition. \ The present behavior
indicates that\ the low temperature state is not a homogeneous ferroelectric
state but, rather, a mesoscopic one, presumably made of large polar
regions. \ At $T_{c}=\allowbreak 413$K, the finite size of the polar
regions\ is $\sim $79\AA\ (19 unit cells)\ and $\sim $114\AA\ (28 unit
cells) respectively in the $[001]$ and $[111]$ directions. \ The longer
correlation length or effective size of the polar regions in the $[111]$
direction is found to be $\xi _{\left[ 111\right] }$ $\sim 1.2\xi _{\left[
001\right] }$ (see dotted line in Fig.3b), consistent with the larger Pb 
displacement component in the $%
[111]$ direction. \ \ This result suggests that the polar regions are in the
form of ellipsoids, preferentially oriented in $[111]$ directions.\ \
Another noticeable difference between the two directions is the appearance
of two kinks in the $\xi _{\lbrack 111]}$ curve, at $\sim $ 450K and $\sim $
413K, that are not visible in the $\xi _{\lbrack 001]}$ curve. \ As
discussed above, the first one, at 450K, reveals the relatively sudden
formation of permanent polar regions, and the second one corresponds to the
transition at $T_{c}$. \ The absence of these kinks in the $\xi _{\lbrack
001]}$ curve\ indicates that the growth of the correlation length in that
direction is only indirectly affected by the formation of the polar regions
at 450K and by their collective realignment at the transition at 413K. \
This result is also consistent with the primary net displacements of Pb and 
three O atoms in one of eight possible $[111]$ directions.

In conclusion, thermal diffuse neutron scattering measurements in the
paraelectric and relaxor\ temperature regions of the PZN crystal clearly
reveal the relatively abrupt formation of polar nanoregions at 450K,
approximately 40 degrees above the transition at $T_{c}=\allowbreak 413$K. \
With decreasing temperature, these polar regions grow and so does their
volume fraction. \ The correlation lengths in the two directions studied are
found to be in the ratio $\xi _{\left[ 111\right] }\sim 1.2\xi _{\left[ 001%
\right] }$ over the whole temperature range, except for two kinks in the $%
\xi _{\left[ 111\right] }$ curve at 450K and 413K. \ The fact that the
correlation lengths remain high below the transition indicates that, at low
temperature, the polar regions subsist at low temperature in what can be
called a mesoscopic ferroelectric phase. Diffuse scattering results at other
reciprocal lattice points are presently being analyzed and will be reported
in a subsequent paper.

We wish to thank R.K. Pattnaik and G. Shirane for helpful discussions. \
This work was supported by DOE Grant No. DE-FG02-00ER45842 and ONR Grant No
N00014-99-1-0738 as well as the Oak Ridge National Laboratory at the High
Flux Isotope Reactor. \


\begin{references}
\bibitem{Yokomizo et al_1278_1970}  Y.Yokomizo, T. Takahashi, and S.
Nomura, J. Phys. Soc. Jpn.\ {\bf 28}, 1278 (1970).

\bibitem{Randall_Bhalla_327_1990}  C.A. Randall and A.S. Bhalla, Jpn. J.
Appl. Phys. {\bf 29}, 327 (1990).

\bibitem{Smolensky_26_1970}  G.A. Smolensky, J. Phys. Soc. Jpn. {\bf 28},
Suppl. 26 (1970).

\bibitem{Cross_241_1987}  L.E. Cross, Ferroelectrics {\bf 76}, 241 (1987).

\bibitem{Khuchua et al_194_1968}  N.P. Khuchua, V.A. Bokov, and I.E. Myl\'{}nikova, 
Sov. Phys.-Solid State. {\bf 10}, 194 (1968).

\bibitem{Kuwata et al_863_1979}  J. Kuwata, K. Uchino,and S. Nomura,
Ferroelectrics {\bf 22}, 863 (1979).

\bibitem{Mulvihill et al_1462_1997}  M. Mulvihill, L.E. Cross, W. Cao, and
K. Uchino, J. Am. Ceram. Soc. {\bf 80}, 1462 (1997).

\bibitem{Kuwata et al_1298_1982}  J. Kuwata, K. Uchino, and S. Nomura, Jpn.
J. Appl. Phys., Part1 {\bf 21}, 1298 (1982).

\bibitem{Liu et al_2810_1999}  S-F. Liu, S-E. Park, T.R. Shrout, and L.E.
Cross, J. Appl. Phys. {\bf 85}, 2810 (1999).

\bibitem{Gehring et al_5216_2000}  P.M. Gehring, S.-E. Park, and G. Shirane,
Phys. Rev. Lett. {\bf 84}, 5216 (2000).

\bibitem{Iwase et al_1419_1999}  T. Iwase, H. Tazawa, K. Fujishiro, Y. Uesu,
and Y. Yamada, J. Phys. Chem. Solids. {\bf 60}, 1419 (1999).

\bibitem{Takesue et al_2001}  N. Takesue, Y. Fujii,and H. You, (unpublished).

\bibitem{Yong et al_14736_2000}  G. Yong, J. Toulouse, R. Erwin, S. Shapiro,
and B. Hennion, Phys. Rev. B {\bf 62}, 14736 (2000).

\bibitem{Egami et al_S935_1998}  T. Egami, S. Teslic, W. Dmowski, P.K.
Davies, and I.-W. Chen, J. Korean Phys. Soc. {\bf 32}, S935 (1998).

\bibitem{Krivoglaz_NewYork_1969}  M.A. Krivoglaz, Theory of X-Ray and
Thermal-Neutron Scattering by Real Crystals, Plenum Press, New York (1969).

\bibitem{Schmatz_NewYork_London_1973}  W. Schmatz, in X-ray and Neutron
Scattering Studies on Disordered Crystals (Edited by Herbert Herman), p.105,
New York and London (1973).

\bibitem{Zhang et al_96_2000}  L. Zhang, M. Dong and Z.-G. Ye, Mater. Sci.
Eng. B. {\bf 78}, 96 (2000).

\bibitem{Gosula et al_221_2000}  V. Gosula, A. Tkachuk, K. Chung, and H.
Chen, J. Phys. Chem. Solids. {\bf 61}, 221 (2000).

\bibitem{Stanley_1971}  H.E. Stanley, Introduction to Phase Transitions and
Critical Phenomena, Oxford University Press, (1971).

\bibitem{Nomura et al_39_1999}  K. Nomura, T. Shingai, S.Ishino, H.
Terauchi, N. Yasuda, and H. Ohwa, J. Phys. Soc. Jpn.\ {\bf 68}, 39 (1999).

\bibitem{Jona_Shirane_NewYorK_1962}  F. Jona. and G. Shirane, Ferroelectric
Crystals, Dover Publications, New York (1962).
\end{references}
\end{document}